\documentclass[12pt]{article}
\usepackage{a4wide}

\usepackage{amssymb,amsmath}
\allowdisplaybreaks

\usepackage{graphicx}
\graphicspath{{fig/}{axodraw/}}

\def\Li(#1,#2){{\rm Li}_{#1}(#2)}
\def\Fc{{\cal F}}
\def\Oc{{\cal O}}
\def\dd{{\rm d}}
\def\MSbar{{\ensuremath{\overline{\rm MS}}}}
\def\QM{\left(\frac{Q^2}{M^2}\right)}
\def\LL{{\rm LL}}
\def\NLL{{\rm NLL}}
\def\NNLL{{\rm NNLL}}
\def\NNNLL{{\rm N^3LL}}
\def\bare{{\rm bare}}

\def\Re{{\rm Re}}

\def\weak{{\rm weak}}

\begin{document}

\setcounter{page}{0}
\thispagestyle{empty}
\rightline{TTP03-06}
\rightline{DESY 03-021}
\rightline{hep-ph/0303016}
\vspace{20mm}

\begin{center}
\textbf{\Large
Fermionic and Scalar Corrections \\[1ex]
for the Abelian Form Factor at Two Loops}
\end{center}

\vspace{5mm}

\begin{center}
{\large B.~Feucht$^{\rm a, *}$, J.~H.~K\"uhn$^{\rm a}$ and S.~Moch$^{\rm b}$}%
\\[4ex]

\textit{$^a$ Institut f\"{u}r Theoretische Teilchenphysik, \\
   Universit\"{a}t Karlsruhe, 76128 Karlsruhe, Germany}\\[2ex]

\textit{$^b$  Deutsches Elektronen-Synchrotron, DESY, \\
        Platanenallee 6, 15738 Zeuthen, Germany}
\end{center}
{\stepcounter{footnote}\renewcommand{\thefootnote}{\fnsymbol{footnote}}%
\footnotetext{B.~Jantzen in later publications}}

\vspace{15mm}

\begin{abstract}
  Two-loop corrections for the form factor in a massive Abelian theory are
  evaluated, which result from the insertion of massless fermion or scalar loops
  into the massive gauge boson propagator.
  The result is valid for arbitrary energies and gauge boson mass.
  Power-suppressed terms vanish rapidly in the high energy region where the
  result is well approximated by a polynomial of third order in
  $\ln(s/M^2)$.
  The relative importance of subleading logarithms is emphasised.
  
  \vspace{5mm}
  
\end{abstract}

\newpage
\section{Introduction\label{sec:introduction}}

One of the central tasks of the next generation of high energy colliders
-- the Large Hadron Collider presently under construction or projects under
consideration like TESLA --
will be the exploration of electroweak interactions at ultrahigh energies:
in the TeV region or beyond.
The control of radiative corrections plays an essential role in this context.
Important differences arise in their structure when comparing low
energies, say up to $\Oc(M_{W,Z})$, with this ultrahigh energy region.
In the first case gauge boson self energies play the dominant role, with
contributions from virtual top quarks or Higgs bosons as most prominent
examples.
At high energies, however, large logarithms arising from virtual gauge boson
exchange become increasingly important.
The leading terms of order $(\alpha_\weak/4\pi)^n \ln^{2n}(s/M^2)$,
often denoted Sudakov logarithms, could easily affect the cross section by 10\%
or more once the energy reaches one or two TeV.
In principle these and even subleading terms have to be summed to all orders.
In practice, in view of the smallness of the weak coupling, it is often
sufficient to restrict the discussion to terms up to order $\alpha_\weak^2$.
(For recent discussions of electroweak Sudakov logarithms see e.g.
\cite{Ciafaloni:1998xg,Hori:2000tm,Denner:2000jv,Denner:2003wi}.)

At present, not only leading logarithms have been
evaluated~\cite{Kuhn:1999de,Fadin:1999bq},
using arguments originally developed in the context of QCD,
next-to-leading (NLL, \cite{Kuhn:1999nn,Kuhn:2000hx})
and even next-to-next-to-leading
(NNLL, \cite{Kuhn:2001hz})
logarithmic corrections have been calculated for four-fermion processes.
Numerically these subleading terms are large and mostly of alternating sign.
To arrive at reliable predictions, say at the level of $\Oc$(1\%),
the evaluation of all two-loop non-power-suppressed corrections for
four-fermion processes seems desirable.
Furthermore it seems useful to test the basic assumption of this approach
that power-suppressed terms of order $M^2/s$ can be neglected.

At the moment this program is completely out of reach, as far as four-fermion
processes in the complete electroweak theory are concerned,
and even in the context of a simplified version like a spontaneously broken
$SU(2)$ gauge theory or an Abelian theory with a massive gauge boson this seems
like an extremely difficult task.
In this present paper we therefore consider a simpler two-loop problem which
nevertheless encompasses already many aspects of the complete calculation:
The contributions from loops of massless fermions or scalars to the vertex
function in an Abelian theory with a massive gauge boson,
which already exhibits many features characteristic for the
complete problem.

The result will be presented for arbitrary $s/M^2$.
This allows to investigate the complete series in the logarithmic expansion
as well as power-suppressed terms.
In the next section we briefly recall the results for the form factor obtained
in \cite{Kuhn:1999nn,Kuhn:2000hx,Kuhn:2001hz}
and introduce our notation.
In section~\ref{sec:nfns} we present the main results of this paper
and discuss its implications.
Section~\ref{sec:summary} contains our summary and conclusions.

\section{The Abelian form factor\label{sec:formfactor}}
Let us begin with a discussion of the form factor for the vector current in
an Abelian gauge theory with a massive gauge boson and massless fermions
in the Sudakov limit.
In Born approximation, one writes 
\begin{equation}
  \Fc_B = \bar\psi(p_2)\gamma_\mu\psi(p_1) ,
\end{equation}
and we study the limit
$s=(p_1-p_2)^2 \to-\infty$ with
on-shell massless fermions, $p_1^2=p_2^2=0$, and
massive gauge bosons, $M^2\ll -s$.
For convenience we choose
$p_{1,2} = (Q/2,0,0,\mp Q/2)$ so that  $2 p_1\!\cdot\! p_2 = Q^2=-s$
and limit the discussion to the spacelike region.
The transition to timelike momentum transfer is easily accomplished through
analytic continuation.

The large logarithmic corrections in the Sudakov limit can be resummed
to all orders of perturbation
theory~\cite{Sen:1981sd,Korchemsky:1989pn}, 
such that the asymptotic behaviour of the form factor is obtained from
\begin{equation}
\label{eq:evolsolf}
  \Fc = F_0(\alpha(M^2))
    \exp \left\{\int_{M^2}^{Q^2}\frac{\dd x}{x}
    \left[\int_{M^2}^{x}\frac{\dd x'}{x'}\gamma(\alpha(x'))+\zeta(\alpha(x))
    +\xi(\alpha(M^2))\right]\right\} \Fc_B .
\end{equation}

The next-to-next-to-leading logarithmic corrections include all the terms of the 
form $\alpha^n\log^{2n-m}(Q^2/M^2)$ with $m=0,~1,~2$. 
To this accuracy, one needs for the anomalous 
dimensions $\gamma$,  $\zeta$ and $\xi$ in Eq.~(\ref{eq:evolsolf}) 
the one-loop results 
\begin{equation}
\label{eq:oneloopadim}
  \gamma^{(1)} = -2C_F \,,\quad
  \zeta^{(1)} = 3C_F\,,\quad
  \xi^{(1)} = 0 \,,\quad
  F_0^{(1)} = - C_F \left( \frac{2}{3}\pi^2 + \frac{7}{2} \right) ,
\end{equation}
as well as the two-loop result for 
the anomalous dimension $\gamma^{(2)}$.
Here we define 
$\gamma = \sum_n \left(\frac{\alpha}{4\pi}\right)^n \gamma^{(n)}$
and similarly for $\zeta$, $\xi$ and $F_0$.
An efficient strategy for the evaluation of the anomalous dimensions
listed in Eq.~(\ref{eq:oneloopadim}) which was
based on the expansion by regions has been described in \cite{Kuhn:1999nn}.
In the \MSbar-scheme, 
including $n_f$ light fermions and $n_s$ light scalars 
in the fundamental representation, $\gamma^{(2)}$
reads~\cite{Kodaira:1982nh,Davies:1984hs,Korchemsky:1987wg}
\begin{equation}
\label{eq:gamma2}
  \gamma^{(2)} = - 2 C_F \left[
    \left( - \frac{\pi^2}{3} + \frac{67}{9} \right) C_A 
    - \frac{20}{9} T_F n_f 
    - \frac{8}{9} T_F n_s
   \right] .
\end{equation}

The result for the form factor, up to $\Oc(\alpha^2)$,
is written in the following form:
\begin{equation}
  \Fc = \Fc_B + \frac{\alpha}{4\pi} \, \Fc^{(1)}
    + \left(\frac{\alpha}{4\pi}\right)^2 \, \Fc^{(2)}
\end{equation}
with
\begin{equation}
  \Fc^{(1)} = \left\{
    \frac{1}{2}\gamma^{(1)} \ln^2\QM
    + \left(\zeta^{(1)} + \xi^{(1)}\right) \ln\QM
    + F_0^{(1)}
    \right\} \Fc_B
\end{equation}
and $\alpha = \alpha_\MSbar(M^2)$.
To obtain $\Fc^{(2)}$, one expands Eq.~(\ref{eq:evolsolf}) for the form
factor at two loops up to next-to-next-to-next-to-leading logarithmic (N$^3$LL)
accuracy,
\begin{align}
\label{eq:llf}
  \Fc^{(2)}_\LL &=
    \frac{1}{8} \big(\gamma^{(1)}\big)^2 \ln^4\QM \Fc_B ,
  \\
\label{eq:nllf}
  \Fc^{(2)}_\NLL &=
    \frac{1}{2}\left(\zeta^{(1)} - \frac{1}{3}\beta_0\right) \gamma^{(1)}
    \ln^3\QM \Fc_B ,
  \\
\label{eq:nnllf}
  \Fc^{(2)}_\NNLL &=
    \frac{1}{2}\biggl(\gamma^{(2)} + \big(\zeta^{(1)}-\beta_0\big)\zeta^{(1)}
      + F_0^{(1)}\gamma^{(1)}\biggr)
    \ln^2\QM \Fc_B ,
  \\
\label{eq:nnnllf}
  \Fc^{(2)}_\NNNLL &=
    \left(\zeta^{(2)} + \xi^{(2)} + F_0^{(1)}\zeta^{(1)}\right)
    \ln\QM \Fc_B ,
\end{align}
where $\xi^{(1)} = 0$ has been omitted. 

Employing the results of Eqs.~(\ref{eq:oneloopadim}) and (\ref{eq:gamma2}),
we see a particular pattern of growing coefficients of the logarithms
which reflects the general structure of logarithmically enhanced
electroweak corrections.
For an Abelian theory ($C_F = 1$, $T_F = 1$, $C_A = 0$),
\begin{align}
\label{eq:abelianF1}
  \Fc^{(1)} &= \biggl\{
    - \ln^2\QM
    + 3 \ln\QM
    - \left( \frac{2}{3} \pi^2 + \frac{7}{2} \right)
    \biggr\} \Fc_B
    \, ,
  \\
\label{eq:abelianF2}
  \Fc^{(2)} &= \biggl\{
    \frac{1}{2} \ln^4\QM
    - \left( \frac{4}{9} n_f + \frac{1}{9} n_s + 3 \right) \ln^3\QM
\nonumber \\ & \qquad\qquad\qquad {}
    + \left( \frac{38}{9} n_f + \frac{25}{18} n_s + \frac{2}{3}\pi^2 + 8 \right)
      \ln^2\QM
    \biggr\} \Fc_B
\end{align}
to NNLL accuracy.
The relatively small coefficient of the leading logarithm and the 
large coefficient of the NNLL term in the 
form factor are clearly indicative of the importance of subleading
logarithmic corrections. 
At N$^3$LL accuracy, the still unknown quantities $\zeta^{(2)}$ and
$\xi^{(2)}$ enter.

\section{Fermionic and scalar contributions at two loops\label{sec:nfns}}

As stated in the introduction, the evaluation of all 
two-loop terms linear in the logarithm or even 
of the two-loop constant terms is desirable.
As a first step, and as a new result, the corrections due to
$n_f$ massless fermions and $n_s$ charged massless scalars 
have been calculated for the Abelian form factor. 
These fermionic and scalar corrections are separately gauge invariant and
renormalisable.
In two loops, they give contributions proportional to $n_f$ and $n_s$
respectively.

\begin{figure}[!bt]
  \centering
  \includegraphics{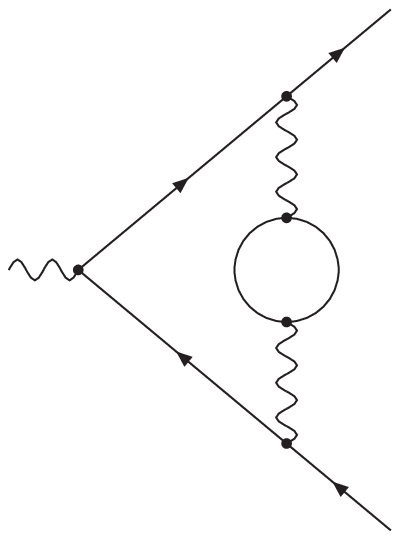}
  \hspace{2cm}
  \includegraphics{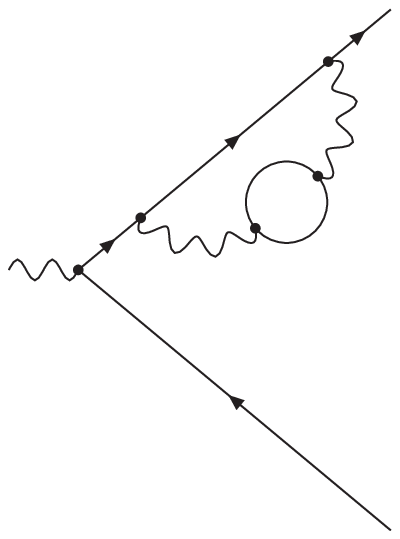}
\caption{%
  Two-loop diagrams contributing to the fermionic and scalar form factor.
  The circle represents a fermion loop for the $n_f$-part or a
  scalar loop for the $n_s$-part of the form factor.}
\label{fig:feynvertex}
\end{figure}

To calculate these terms, the two-loop diagrams depicted in
Fig.~\ref{fig:feynvertex}
and the gauge boson mass renormalisation
have been evaluated.
This calculation was done using dimensional regularisation.
Due to the simple topologies of the diagrams in Fig.~\ref{fig:feynvertex},
the integration corresponding to the insertion of the fermion or scalar loop
can be done first, leaving only one-loop integrals with an
additional massless propagator of non-integer power.
Tensor integrals were reduced to scalar ones by employing the method developed
by Passarino and Veltman~\cite{Passarino:1979jh}.
After using partial integration for further reduction, we were left with only
three master integrals which could be calculated with the help of Feynman
parameters.
Alternatively, it is also possible to evaluate all integrals
resulting from the tensor reduction directly with the method
of nested sums~\cite{Vermaseren:1998uu,Moch:2001zr,Moch:2002hm}.

The renormalisation of the coupling~$\alpha$ was performed in the \MSbar-scheme
at the scale $\mu^2 = M^2$.
The ``on-shell mass''~$M$ of the gauge boson is defined as the location of the
zero of the real part of the inverse propagator,
\begin{equation}
  \Re\left[ p^2 - M_\bare^2 - \Sigma_\bare(p^2) \right]_{p^2 = M^2} = 0 ,
\end{equation}
thus
\begin{equation}
  M_\bare^2 = M^2 - \Re\,\Sigma_\bare(M^2) .
\end{equation}
The difference between this definition and another one, where $M^2$ is defined
as the real part of the location of the pole of the propagator,
becomes relevant only in higher orders.

The contributions of $n_f$ fermions and $n_s$ scalars,
together with the Born term and the one-loop result,
yield
\begin{align}
\label{eq:FFcomplete}
  \Fc 
  &= \Biggl\{
    1
    + \frac{\alpha}{4\pi} \Biggl[
      - (1-z)^2 \biggl(
        2\,\Li(2,1-z)
        + \ln^2(z) 
        + \frac{\pi^2}{3}
        \biggr)
      - (3-2z) \ln(z)
      - \frac{7}{2} + 2z
      \Biggr]
\nonumber \\& \qquad {}
    + n_f \left(\frac{\alpha}{4\pi}\right)^2 \Biggl[
      (1-4z+3z^2) \biggl(
        \frac{8}{3}\,\Li(3,z)
        + \frac{8}{3} \ln(z)\,\Li(2,1-z)
        + \frac{4}{9} \ln^3(z)
\nonumber \\& \qquad \qquad\qquad\qquad \qquad\qquad\qquad\quad {}
        + \frac{4}{3} \ln^2(z) \ln(1-z)
        - \frac{4}{9} \pi^2 \ln(z)
        \biggr)
\nonumber \\& \qquad \qquad\qquad\qquad
      + (1-z)^2 \biggl(
        \frac{16}{9}\,\Li(2,1-z)
        + \frac{8}{27} \pi^2
        \biggr)
      + \biggl( \frac{38}{9} - \frac{52}{9}z + \frac{8}{9}z^2 \biggr) \ln^2(z)
\nonumber \\& \qquad \qquad\qquad\qquad
      + \biggl( \frac{34}{3} - \frac{88}{9}z \biggr) \ln(z)
      + \frac{115}{9} - \frac{88}{9}z
      \Biggr]
\nonumber \\& \qquad {}
    + n_s \left(\frac{\alpha}{4\pi}\right)^2 \Biggl[
      (1-4z+3z^2) \biggl(
        \frac{2}{3}\,\Li(3,z)
        + \frac{2}{3} \ln(z)\,\Li(2,1-z) 
        + \frac{1}{9} \ln^3(z) 
\nonumber \\& \qquad \qquad\qquad\qquad \qquad\qquad\qquad\quad {}
        + \frac{1}{3} \ln^2(z) \ln(1-z) 
        - \frac{\pi^2}{9} \ln(z)
        \biggr)
\nonumber \\& \qquad \qquad\qquad\qquad {}
      + (1-z)^2 \biggl(
        \frac{10}{9}\,\Li(2,1-z)
        + \frac{5}{27} \pi^2
        \biggr)
      + \biggl( \frac{25}{18} - \frac{19}{9}z + \frac{5}{9}z^2 \biggr) \ln^2(z)
\nonumber \\& \qquad \qquad\qquad\qquad {}
      + \biggl( \frac{23}{6} - \frac{28}{9} z \biggr) \ln(z)
      + \frac{157}{36}
      - \frac{28}{9}z
      \Biggr]
    \Biggr\} \,
    \Fc_B \, ,
\end{align}
for arbitrary $z=M^2/Q^2$.
From Eq.~(\ref{eq:FFcomplete})
one easily derives the large logarithms in the high energy limit,
i.e. $z=M^2/Q^2 \to 0$,
\begin{align}
\label{eq:FFlogs}
  \Fc 
  &= \Biggl\{
    1 
    + \frac{\alpha}{4 \pi} \Biggl[
      - \ln^2 (z)
      - 3 \ln(z)
      - \frac{2}{3} \pi^2
      - \frac{7}{2}
      \Biggr]
\nonumber \\& \qquad {}
    + n_f \left(\frac{\alpha}{4\pi}\right)^2 \Biggl[
      \frac{4}{9} \ln^3(z)
      + \frac{38}{9} \ln^2(z)
      + \frac{34}{3} \ln(z)
      + \frac{16}{27} \pi^2
      + \frac{115}{9}
      \Biggr]
\nonumber \\& \qquad {}
    + n_s \left(\frac{\alpha}{4\pi}\right)^2 \Bigg[
      \frac{1}{9} \ln^3(z)
      + \frac{25}{18} \ln^2(z)
      + \frac{23}{6} \ln(z)
      + \frac{10}{27} \pi^2
      + \frac{157}{36}
      \Biggr]
   \Biggr\} \,
   \Fc_B \, .
\end{align}
This result has also been obtained through dispersion relations similarly to the
technique discussed in \cite{Kniehl:1988id}.

The terms up to $\ln^2(z)$ at two loops
agree with the result from the evolution equation, 
i.e. Eqs.~(\ref{eq:llf})--(\ref{eq:nnllf}) and (\ref{eq:abelianF2}).
The coefficients of the terms proportional to $\ln(z)$
and the constant terms represent a new result.
With the help of Eq.~(\ref{eq:nnnllf}), one can determine the
contributions of light fermions and light charged scalars to the sum
of the anomalous dimensions $\zeta^{(2)}$ and $\xi^{(2)}$,
\begin{equation}
  \zeta^{(2)} + \xi^{(2)} \Biggl|_{n_f+n_s} =
    - C_F \biggl[
      \frac{34}{3} T_F n_f 
      + \frac{23}{6} T_F n_s
   \biggr] .
\end{equation}

\begin{figure}[!bt]
  \centering
  \includegraphics{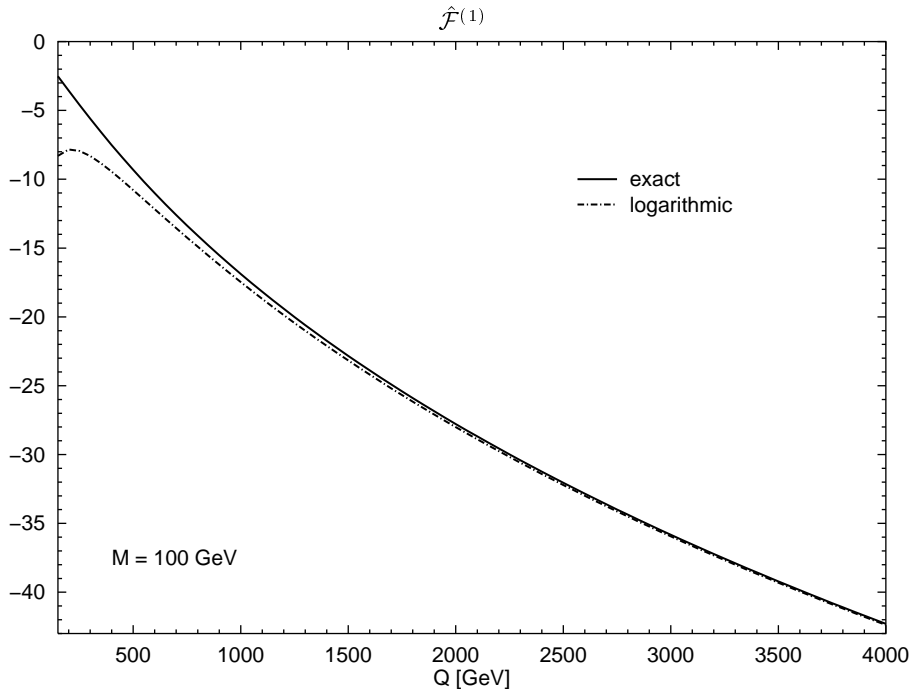}
\caption{%
  The one-loop contribution $\hat\Fc^{(1)}$
  of the Abelian form factor as defined in Eq.~(\ref{eq:FFnfscaling}).
  Plotted are the exact result~(\ref{eq:F1complete})
  and the complete logarithmic approximation~(\ref{eq:F1logs}).}
\label{fig:F1complete}
\end{figure}

\begin{figure}[!bt]
  \centering
  \includegraphics{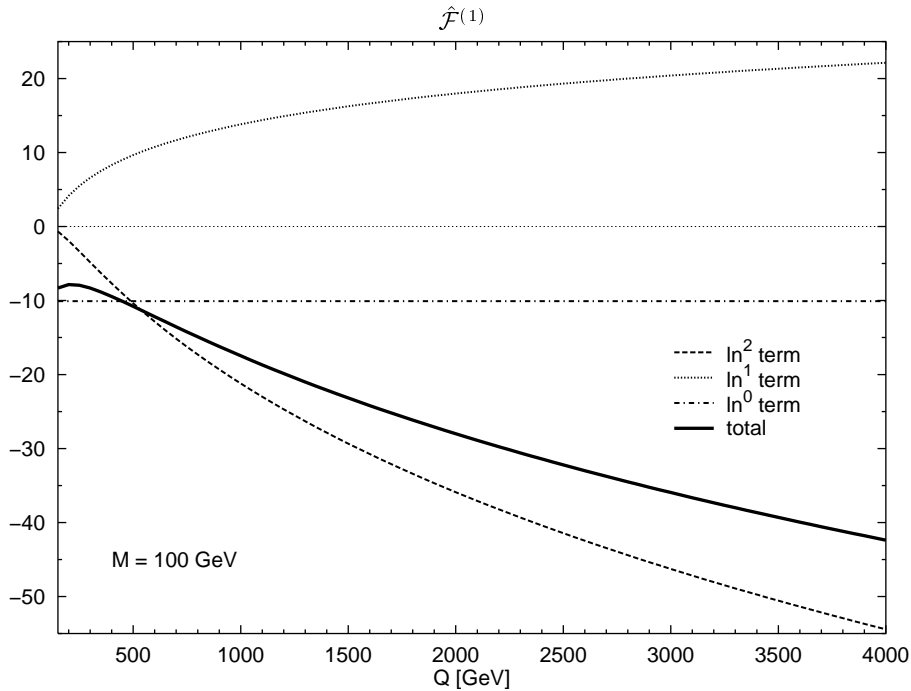}
\caption{%
  The one-loop contribution $\hat\Fc^{(1)}$
  of the Abelian form factor as defined in Eq.~(\ref{eq:FFnfscaling}).
  Plotted are the individual contributions of the large logarithms
  as well as the complete logarithmic approximation.}
\label{fig:F1logs}
\end{figure}

Let us define the form factor in terms of scaling functions as 
\begin{equation}
\label{eq:FFnfscaling}
  \Fc = \left\{
    1 
    + \frac{\alpha}{4\pi} \hat\Fc^{(1)}
    + \left(\frac{\alpha}{4\pi}\right)^2 \left(
      n_f \hat\Fc^{(2)}_{n_f}
      + n_s \hat\Fc^{(2)}_{n_s}
      \right)
    \right\}
    \Fc_B \, .
\end{equation}
Before entering the discussion of the two-loop result,
let us recapitulate some qualitative features of the one-loop result
which will reappear for the two-loop integral.
The result for $\hat\Fc^{(1)}$ has been known since long
(e.g.~\cite{Bohm:1986rj,Grzadkowski:1987pm}),
\begin{align}
\label{eq:F1complete}
  \hat\Fc^{(1)}
  &=
      - (1-z)^2 \biggl(
        2\,\Li(2,1-z)
        + \ln^2(z) 
        + \frac{\pi^2}{3}
        \biggr)
      - (3-2z) \ln(z)
      - \frac{7}{2} + 2z
  \\
\label{eq:F1logs}
  &\xrightarrow{z\to 0}
      - \ln^2\left(\frac{1}{z}\right)
      + 3 \ln\left(\frac{1}{z}\right)
      - \left( \frac{2}{3} \pi^2 + \frac{7}{2} \right)
      .
\end{align}
Let us adopt $M = 100$~GeV as a typical choice for electroweak gauge boson
masses and compare the complete answer~(\ref{eq:F1complete})
with the approximation where power-suppressed terms are
neglected (Eq.~\ref{eq:F1logs}).
For energies above 500~GeV, corresponding to $z \approx 0.04$,
power-suppressed terms are small,
and above 1000~GeV they can safely be neglected (Fig.~\ref{fig:F1complete}).
On the other hand, leading, subleading logarithm and constant are of
alternating sign, the respective coefficient increases markedly and large
compensations are apparent even in the multi-TeV region
(Fig.~\ref{fig:F1logs}).

We will now turn to the two-loop result, Eqs.~(\ref{eq:FFcomplete})
and~(\ref{eq:FFlogs}).
To assess the quality of the logarithmic approximation of
Eq.~(\ref{eq:FFlogs}), we plot
$\hat\Fc^{(2)}_{n_f}$ and $\hat\Fc^{(2)}_{n_s}$
as defined in Eq.~(\ref{eq:FFnfscaling})
for $M=100$~GeV. 
\begin{figure}[p]
  \centering
  \includegraphics{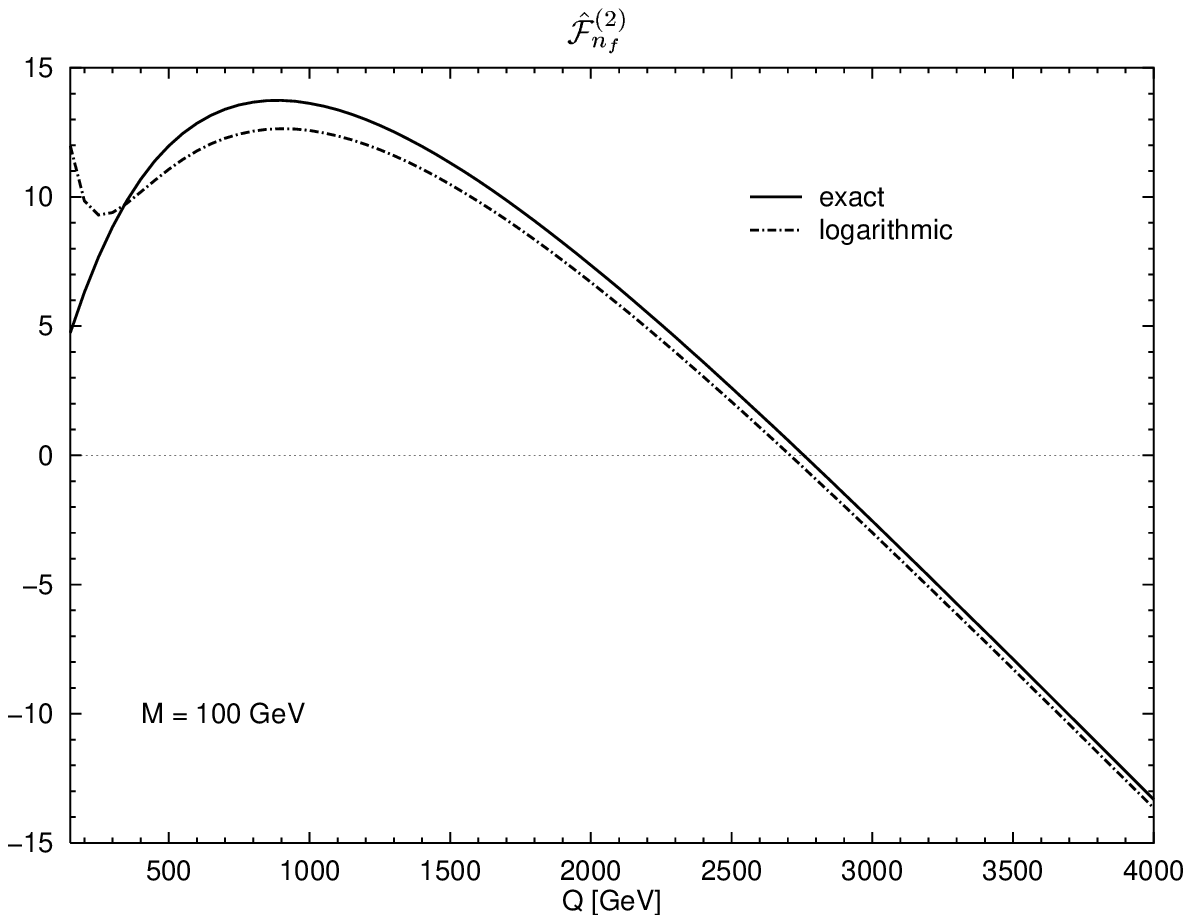}\\[1cm]
  \includegraphics{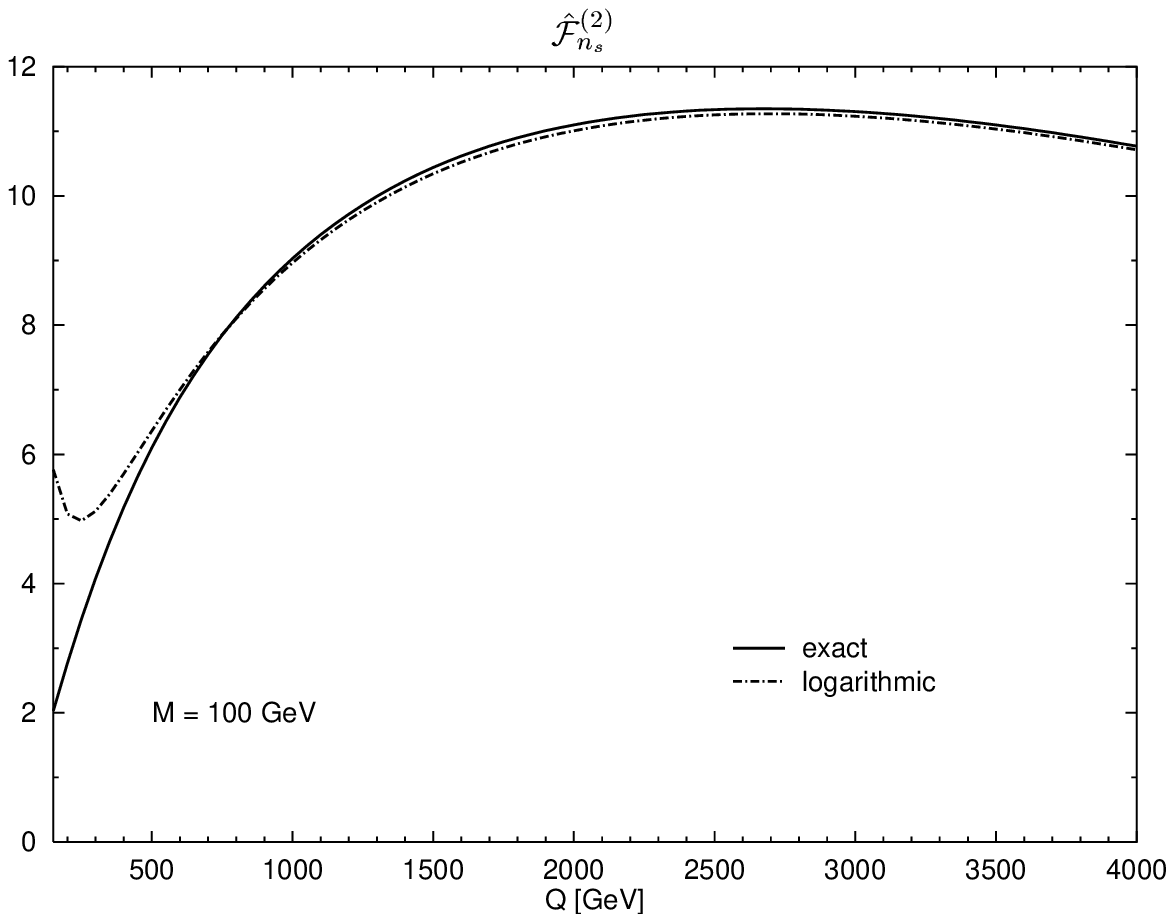}
\caption{%
  The fermionic contribution $\hat\Fc^{(2)}_{n_f}$
  and the scalar contribution $\hat\Fc^{(2)}_{n_s}$
  of the Abelian form factor at two loops as defined in
  Eq.~(\ref{eq:FFnfscaling}). 
  Plotted are the exact result and the complete logarithmic approximation.}
\label{fig:FFcomplete}
\end{figure}%
In Fig.~\ref{fig:FFcomplete} the exact result for
$\hat\Fc^{(2)}_{n_f}$ and $\hat\Fc^{(2)}_{n_s}$
as given in Eq.~(\ref{eq:FFcomplete}) is compared
with the complete logarithmic approximation from Eq.~(\ref{eq:FFlogs}).
Good agreement is observed over a wide range in $Q$.
Power-suppressed corrections proportional to $M^2/Q^2$ are small
in the high energy regime $Q \gtrsim 500$~GeV
relevant for future colliders.
This result justifies the approximation which neglects power-suppressed terms
in the calculation of the whole form factor.

In the next step we investigate the quality of leading, next-to-leading
and next-to-next-to-leading logarithmic approximations.
\begin{figure}[p]
  \centering
  \includegraphics{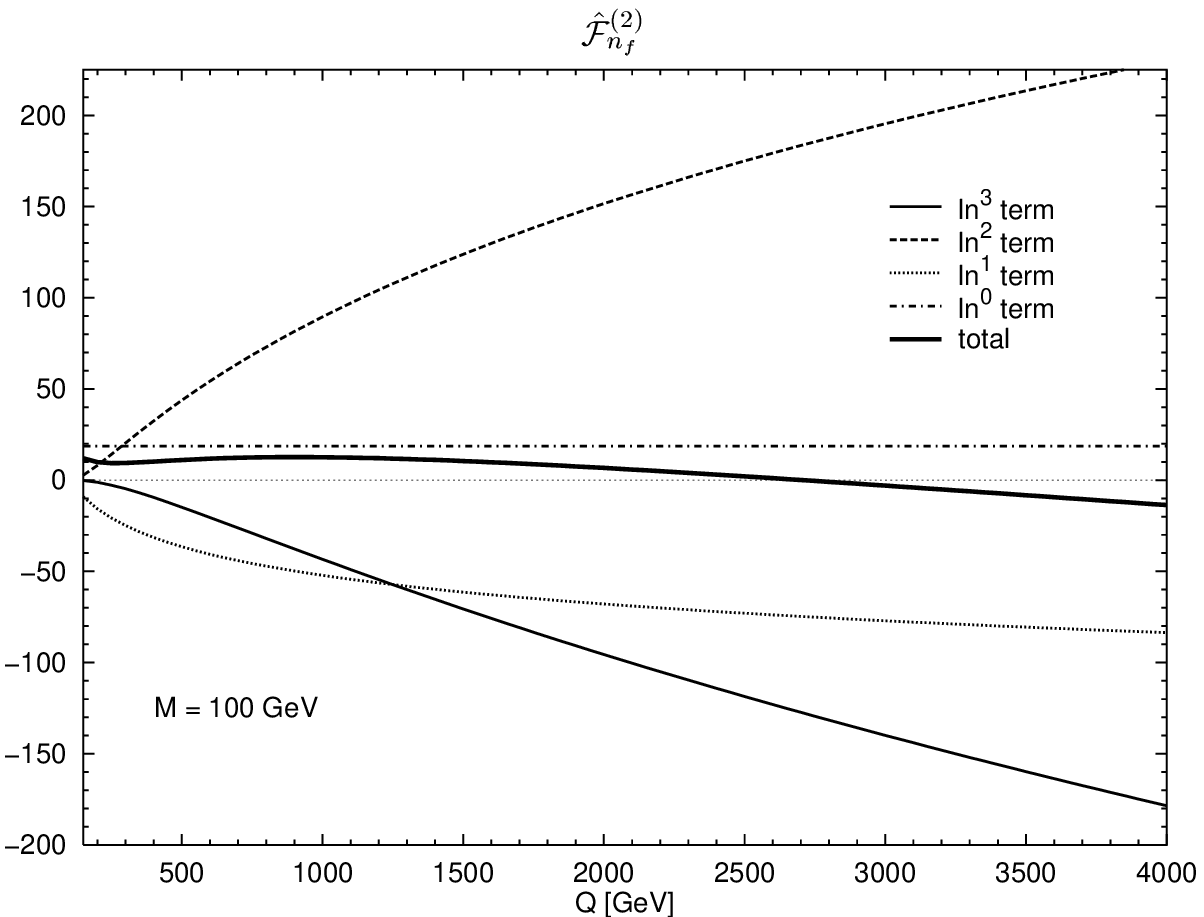}\\[1cm]
  \includegraphics{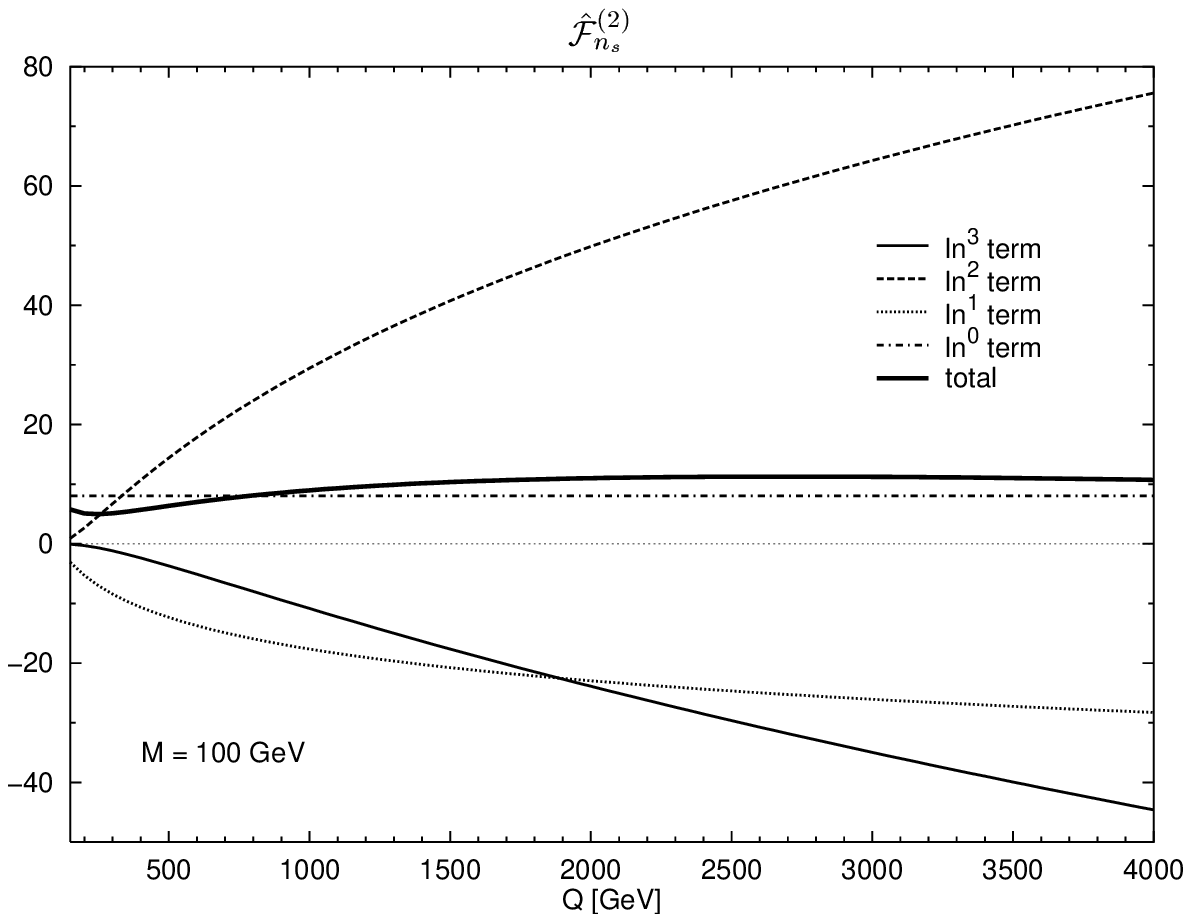}
\caption{%
  The fermionic contribution $\hat\Fc^{(2)}_{n_f}$
  and the scalar contribution $\hat\Fc^{(2)}_{n_s}$
  of the Abelian form factor at two loops as defined in
  Eq.~(\ref{eq:FFnfscaling}). 
  Plotted are the individual contributions of the large logarithms
  as well as the complete logarithmic approximation.}
\label{fig:FFlogs}
\end{figure}%
A pattern of growing coefficients of the logarithms
in Eq.~(\ref{eq:FFlogs}) is observed which
continues up to the constant term.
This is evident from Fig.~\ref{fig:FFlogs} which shows
the individual contributions of different powers of logarithms
as given in Eq.~(\ref{eq:FFlogs})
as well as the complete logarithmic approximation.
Large cancellations between subsequent powers of logarithms are observed.
Thus the full result is small compared to the size of the individual
contributions.
Even at 4~TeV the constant term
is comparable in size to the full result.
Considering e.g. the $n_f$-part, the leading logarithm alone would
overshoot the full result by a factor of more than 10.

\section{Summary and conclusions\label{sec:summary}}

Two-loop contributions to the form factor resulting from virtual massless
fermion and scalar loops
have been evaluated in the context of a massive Abelian theory.
It is demonstrated that power-suppressed terms become small for
$M^2/s \lesssim 1/25$.
The leading and subleading logarithmic terms of order
$\ln^3(s/M^2)$ and $\ln^2(s/M^2)$ reproduce those derived
in~\cite{Kuhn:2001hz}.

A marked increase is observed for the coefficients of the linear logarithm and
the constant term.
To arrive at a reliable prediction for electroweak processes in the TeV region,
the control of these subleading terms seems to be required also for the complete
four-fermion process.

\section*{Acknowledgements}

We would like to thank V.~Smirnov and P.~Uwer for instructive
discussions
and A.~Penin and S.~Pozzorini for discussions and for carefully reading the
manuscript.
The work of J.H.K. and S.M. was supported by the DFG under contract FOR~264/2-1,
and by the BMBF under grant BMBF-05HT1VKA/3.

\bibliographystyle{bibspecial}
\bibliography{nfns-paper}

\end{document}